\documentstyle[12pt]{article}

\title{GENERALIZED SIMULATED ANNEALING}
\author{Constantino TSALLIS$^{1,2}$ and Daniel A. STARIOLO$^{1}$ \\
1 - Centro Brasileiro de Pesquisas F\'{\i}sicas
\\ Rua Xavier Sigaud,
150 \\ 22290-180 -- Rio de Janeiro -- RJ, Brazil\\
2 - Department of Physics and Astronomy and\\
Center for Fundamental Materials Research\\
Michigan State University\\
East Lansing, Michigan 48824-1116, USA}
\date{}
\begin{document}
\maketitle

\begin{abstract}

We propose a new stochastic algorithm ({\it generalized simulated
annealing}) for computationally finding the {\it global} minimum of a
given (not necessarily convex) energy/cost function defined in a
continuous D-dimensional space. This algorithm recovers, as
particular cases, the so called {\it classical} (``Boltzmann
machine'') and {\it fast} (``Cauchy machine'') simulated annealings,
and can be quicker than both.

\vspace{3mm}
\end{abstract}

\newpage
\section{INTRODUCTION}

The central step of an enormous variety of problems (in Physics,
Chemistry, Statistics, Neural Networks, Engineering, Economics) is
the minimization of an appropriate energy/cost function defined in a
D-dimensional continuous space $(\vec{x}\in I\!\!\!\!R^D)$. If the
energy is {\it convex} (single minimum), any gradient descent method
easily solves the problem. But if the energy  is {\it nonconvex}
(multiple extrema) the solution requires more  sophisticated methods,
since a gradient descent procedure could easily trap the system in a
{\it local} minimum (instead of one of the {\it global} minima we are
looking for). This sophistication must necessarily involve possible
``hill climbings'' (for detrapping from local minima), and can be
heavily computer-time-consuming. Consequently, various algorithmic
strategies have been developed along the years for making this
important problem increasingly tractable. One of the generically
most efficient (hence popular) methods is {\it simulated annealing}, to
which this paper is dedicated. In this technique, one or more
artificial temperatures are introduced and gradually cooled, in
complete analogy with the well known annealing technique frequently
used in Metallurgy for making a molten metal to reach its crystalline state
({\it global} minimum of the thermodynamical energy). This artificial
temperature (or set of temperatures) acts as a {\it source of
stochasticity}, extremely convenient for eventually detrapping from
local minima. Near the end of the process, the system hopefully is
inside the attractive basin of the {\it global} minimum (or in one of
the global minima, if more than one exists, i.e., if there is {\it
degeneracy}), the temperature is practically zero, and the procedure
asymptotically becomes a gradient descent one. The challenge is to
cool the temperature the quickest we can {\it but} still having the
guarantee that no definite trapping in {\it any} local minimum will
occur. More precisely speaking, we search for the {\it quickest
annealing} (i.e., in some sense approaching a {\it quenching}) which
preserves the probability of ending in a global minimum being equal
to one. The first nontrivial solution along this line was provided in
1983 by Kirkpatrick et al \cite{ki} for classical systems, and was
extended in 1986 by Ceperley and Alder \cite{ce} for quantum
systems. It strictly follows quasi-equilibrium Boltzmann-Gibbs
statistics. The system ``tries'' to visit, according to a {\it
visiting distribution} assumed to be {\it Gaussian} (i.e., a {\it
local} search distribution) in the neighborhood of its actual state
$\vec{x}$. The jump is {\it always accepted} if it is down hill (of
the energy/cost funtion); if it is hill climbing, it {\it might be
accepted} according to an {\it acceptance probability} assumed to be
the canonical-ensemble Boltzmann-Gibbs one. Geman and Geman \cite{ge}
showed, for the classical case, that a necessary and sufficient
condition for having probability one of ending in a global minimum is
that the temperature decreases {\it logarithmically} with time.
This algorithm is sometimes referred to as {\it classical simulated
annealing} (CSA) or {\it Boltzmann machine}. We easily recognize
that, if instead of decreasing, the temperature was maintained fixed,
this procedure precisely is the well known Metropolis et al \cite{me}
one for simulating thermostatistical equilibrium.

The next interesting step along the present line was Szu's 1987
proposal \cite{zu} of using a Cauchy-Lorentz visiting distribution,
instead of the Gaussian one. This is a {\it semi-local} search
distribution: the jumps are frequently local, but can occasionally be
quite long (in fact, this is a L\'evy-flight-like distribution). The
acceptance algorithm remains the same as before. As Szu and Hartley
showed, the
cooling can now be much faster (the temperature is now allowed to
decrease like the {\it inverse} of time), which makes the entire
procedure quite more efficient. This algorithm is referred to as {\it
fast simulated annealing} (FSA) or {\it Cauchy machine}.

The goal of the present work is to generalize {\it both} annealings
within an unified picture which is inspired in the recently
Generalized Statistical Mechanics \cite{sa,cu} (see also \cite{ex,som}),
 with the
supplementary bonus of providing an algorithm which is {\it even
quicker} than that of Szu's. In Section 2, we briefly review this
generalized thermostatistics, describe the optimization algorithm and
prove that, if the cooling rithm is appropriate, the probability of
ending in a global minimum equals one. In Section 3, we numerically
discuss a simple $D=1$ example. Finally, we conclude in Section 4.

\section{GENERALIZED SIMULATED ANNEALING (GSA)}

Inspired by multifractals, one of us proposed \cite{sa} a generalized
entropy $S_q$ as follows
\begin{equation}
S_q=k\ \frac{1-\displaystyle{\sum_{i}}\ p^q_i}{q-1}\;\;\;\;\;\;\;\;\;\;\;\;\;
(q\in I\!\!\!\!R)
\end{equation}
where $\{p_i\}$ are the probabilities of the microscopic configurations
and $k$ is a conventional positive constant. In the $q\rightarrow 1$
limit, $S_q$ recovers the well known Shannon expression
$-k_B\displaystyle{\sum_{i}}\ p_i\ln p_i$. Optimization of this
entropy for the canonical ensemble yields
\begin{equation}
p_i=\frac{[1-\beta (1-q)E_i]^{\frac{1}{1-q}}}{Z_q}
\end{equation}
with
\begin{equation}
Z_q\equiv \sum_{i}\ [1-\beta (1-q)E_i]^{\frac{1}{1-q}}
\end{equation}
where $\beta \equiv 1/kT$ is a Lagrange parameter, and $\{E_i\}$ is
the energy spectrum. We immediately verify that, in the $q\rightarrow
1$ limit,
we recover Boltzmann-Gibbs statistics, namely $p_i=\exp (-\beta
E_i)/Z_1$ with $Z_1\equiv \displaystyle{\sum_{i}}\exp (-\beta E_i)$.

Let us now focus the {\it acceptance probability}
$P_{q_A}(\vec{x}_t\rightarrow \vec{x}_{t+1})$, where $t$ is the
discrete time $(t=1,2,3,\cdots )$ corresponding to the computer
iterations. For the Boltzmann machine $(q_A=1)$ we have, for
example, the Metropolis algorithm \cite{me} :
\begin{equation}
P_1(\vec{x}_t\rightarrow \vec{x}_{t+1})=\left\{
\begin{array}{ll}
1 & \mbox{if} ~E(\vec{x}_{t+1})<E(\vec{x}_t) \\
e^{[E(\vec{x}_t)-E(\vec{x}_{t+1})]/T^A_1(t)}
& \mbox{if} ~E(\vec{x}_{t+1})\geq E(\vec{x}_t)
\end{array}
\right.
\end{equation}
where $T^A_1(t)$ is the $q_A=1$ {\it acceptance temperature} at time
$t$ ($k=1$ from now on). We see that $T^A_1(t)=+0$ implies $P_1=1$ if
$E(\vec{x}_{t+1})<E(\vec{x}_t)$, and $P_1=0$ if
$E(\vec{x}_{t+1}) \geq E(\vec{x}_t)$. These limits are important
for the asymptotic stabilization of the annealing process,and
we will require them to be satisfied by the generalized form.
Eq.(4) satisfies detailed balance.
A generalization of Eq.(4) that also satisfies the detailed balance
condition, and asymptotically generates states distributed
according to Eq.(2), must involve the ratio of terms of the form
$1-\beta (1-q) E$. Nevertheless, we could not find a
generalization along this line which satisfies the $T=0$ limits
mentioned above. Instead, we worked with a form that generalizes
Eq.(4), satisfies the limits at $T=0$, and goes to an
equilibrium distribution, although generically different from
that of Eq.(2). This generalized acceptance probability reads:
\begin{equation}
P_{q_A}(\vec{x}_t\rightarrow \vec{x}_{t+1})=\left\{
\begin{array}{ll}
1 & \mbox{if} ~E(\vec{x}_{t+1})<E(\vec{x}_t) \\
\frac{1}
{\left[1+(q_A-1)(E(\vec{x}_{t+1})-E(\vec{x}_t))/T^A_{q_A}\right]
^{\frac{1}{q_A-1}}} & \mbox{if}~ E(\vec{x}_{t+1})\geq E(\vec{x}_t)
\end{array}
\right.
\end{equation}
Although it is possible to work under generic conditions, for
simplicity we shall assume here that $E(\vec{x})\geq 0\ (\forall
\vec{x})$. Moreover, we shall assume that $q_A\geq 1$, so
$T^A_{q_A}(t)$ can decrease {\it down to zero} without any type of
singularities. Within these hyphothesis,
$P_{q_A}\in [0,1]\ (\forall q_A)$, and, for $T^A_{q_A}(t)$
decreasing from infinity to zero, $P_{q_A}$ monotonically varies from
$1$ to $0$  if $E(\vec{x}_{t+1})\geq E(\vec{x}_t)$,  and
equals 1 whenever $E(\vec{x}_{t+1})<E(\vec{x}_t)$.

We can now focus the $\vec{x}_t\rightarrow \vec{x}_{t+1}$ isotropic
{\it visiting distribution} $g_{q_V}(\Delta x_t)$ where $\Delta
x_t\equiv (\vec{x}_{t+1}-\vec{x}_t)$. It satisfies
\begin{equation}
\Omega_D\int^{\infty}_{0}d\rho \ \rho^{D-1}g_{q_V}(\rho )=1
\end{equation}
where $\Omega_D\equiv D\Pi^{D/2}/\Gamma \left(\frac{D}{2}+1\right)$
is the D-dimensional complete solid angle. For the Boltzmann machine
$(q_V=1)$ we have \cite{ki,zu}
\begin{equation}
g_1(\Delta x_t)\propto \exp \left[-\frac{(\Delta x_t)^2}{T_1^V(t
)}\right]
\end{equation}
where $T^V_1(t)$ is the $q_V=1$ {\it visiting temperature} at time
$t$. Using condition (6) we obtain
\begin{equation}
g_1(\Delta x_t)=\frac{e^{-\frac{(\Delta x_t)^2}{T_1^V(t)}}}
{[\pi \ T_1^V(t)]^{D/2}}
\end{equation}
For the Cauchy machine $(q_V=2)$ we have \cite{zu}
\begin{equation}
g_2(\Delta x_t)\propto \frac{T^V_2(t)}
{\{[T_2^V(t)]^2+(\Delta x_t)^2\}^{\frac{D+1}{2}}}
\end{equation}
where $T_2^V(t)$ is the $q_V=2$ {\it visiting temperature} at time $t$.
The functional form of Eq. (9) is the D-dimensional Fourier
transform of $\exp \{-T^V_2(t)|\vec{y}|\}$ (see \cite{zu}). Using
condition (6) we obtain
\begin{equation}
g_2(\Delta x_t)=\frac{\Gamma (\frac{D+1}{2})}{\pi^{\frac{D+1}{2}}}\
\frac{T^V_2(t)}{\{[T^V_2(t)]^2+(\Delta x_t)^2\}^{\frac{D+1}{2}}}
\end{equation}
Within the present scheme, a natural proposal for unifying (8) and (10)
is
\begin{equation}
g_{q_V}(\Delta x_t)=c\frac{[T^V_{q_V}(t)]^d}
{\{[T^V_{q_V}(t)]^e+(q_V-1)b(\Delta x_t)^2\}^{\frac{a}{q_V-1}}}
\end{equation}
where $a,b,c,d$ and $e$ are $(q_V,D)$-dependent pure numbers to be
determined. Using condition (6) and recalling that $\Delta x_t$ may
carry dimensions (e.g, [length]) we immediately establish that
\begin{equation}
d=e\frac{2a-D(q_V-1)}{2(q_V-1)}\;\;\;\;\;\;\;\;(\forall q_V,\forall D)
\end{equation}
To further decrease the number of independent pure numbers to be
determined, let us address a central point, namely the fact that the
method has to guarantee that, at the $t\rightarrow \infty$ limit, the
system must be at a {\it global} minimum. For this to occur (see
\cite{zu} and references therein) the state visiting must be
{\it ``infinite often in time} $(iot)$'', which indeed occurs if
$\displaystyle{\sum_{t=t_0}^{\infty}}\ g_{q_V}(\Delta x_{t_0})$ {\it
diverges} for fixed $\Delta x_{t_0}$ with $t_0>>1$. Under these
conditions we have that
\begin{equation}
\sum_{t=t_0}^{\infty}\ g_{q_V}(\Delta x_{t_0})\propto
\sum_{t=t_0}^{\infty}[T^V_{q_V}(t)]^d
\end{equation}
We know \cite{zu} that, for arbitrary $D$, $T_1^V(t)=T_1^V(1)\ln
2/\ln (1+t)$ and $T^V_2(t)=T^V_2(1)/t$, which are conveniently unified with
\begin{eqnarray}
T^V_{q_V}(t) &=& T_{q_V}(1)\frac{2^{q_V-1}-1}{(1+t)^{q_V-1}-1} \\
&\sim &T_{q_V}(1)\frac{2^{q_V-1}-1}{t^{q_V-1}}\;\;\;\;\;\;\;\;\;\;
\;\;\;\;\;\;(t\rightarrow
\infty )\nonumber \;\;\;\;\;\;\;\;\;\;\;\;\;\;\;\;\;\;\;\;\;\;\;\;\;
\;\;\;\;\;\;\;\;\;\;\;\;\;\;\;\;\;\;\;\;\;\;\;\;\;\;\;\;\;\;\;\;(14')
\end{eqnarray}
Replacing (14') into Eq. (13) we obtain
\begin{equation}
\sum_{t=t_0}^{\infty}\ g_{q_V}(\Delta x_{t_0})\propto \sum_{t=t_0}^{\infty}
\frac{1}{t^{(q_V-1)d}}
\end{equation}
For arbitrary $D$ and $q_V=1,2$ it is \cite{zu} $(q_V-1)d=1$. We
assume, for simplicity, that the same holds $\forall q_V$, hence
\begin{equation}
d=\frac{1}{q_V-1}\;\;\;\;\;\;\;\;\;\;\;\;(\forall q_V,\forall D)
\end{equation}
consequently the series (15) is the {\it harmonic} one, hence {\it
diverges} (logarithmically) as desired. If we use Eqs. (12) and (16)
into (11) we obtain
\begin{equation}
g_{q_V}(\Delta x_t)=c\frac{[T^V_{q_V}(t)]^{\frac{-D}{2a-D(q_V-1)}}}
{\left\{{
1+(q_V-1)b
\frac{(\Delta x_t)^2}{[T^V_{q_V}(t)]^{\frac{2}{2a-D(q_V-1)}}}
}\right\}
^{\frac{a}{q_V-1}}
}
\end{equation}
For $q_V=1$, Eq. (17) must recover Eq. (8), hence $b=1$ and $a=1$ (for
arbitrary $D$). For $q_V=2$, Eq. (17) must recover Eq. (10), hence
$b=1$ and $a=\frac{D+1}{2}$ (for arbitrary $D$). For simplicity we assume
\begin{equation}
b=1\;\;\;\;\;\;\;\;\;\;(\forall q_V,\ \forall D)
\end{equation}
Finally, condition (6) univocally determines the normalizing pure
number $c$ as a function of the rest of the free parameters. Using
this and Eq. (18) into Eq. (17) yields
\begin{equation}
g_{q_V}(\Delta x_t) = \left(\frac{q_V-1}{\pi}\right)^{D/2}\
\frac{\Gamma \left(\frac{a}{q_V-1}\right)}
{\Gamma\left(\frac{a}{q_V-1}-\frac{D}{2}\right)}\
\frac{[T^V_{q_V}]^{-\frac{D}{2a-D(q_V-1)}}}
{\left\{1+(q_V-1)\frac{(\Delta x_t)^2}{[T^V_{q_V}(t)]
^{\frac{2}{2a-D(q_V-1)}}}\right\}^{\frac{a}{q_V-1}}}
\end{equation}
where {\it only one} undetermined pure number (namely $a(q_V,D)$) is
now left. It satisfies, as already mentioned,  $a(1,D)=1$ and
$a(2,D)=(D+1)/2$. Although more general forms are possible, we  shall
adopt the simplest $q_V$-dependence, namely a linear interpolation, hence
\begin{equation}
a=1+\frac{D-1}{2}\ (q_V-1)\;\;\;\;\;\;\;\;\;\;
(\forall q_V,\ \forall D)
\end{equation}
Replacing this into Eq. (19) we otain our {\it final} visiting distribution
\begin{equation}
g_{q_V}(\Delta x_t)=\left(\frac{q_V-1}{\pi}\right)^{D/2}\
\frac{\Gamma \left(\frac{1}{q_V-1}+\frac{D-1}{2}\right)}
{\Gamma\left(\frac{1}{q_V-1}-\frac{1}{2}\right)}
\ \frac{[T^V_{q_V}(t)]^{-\frac{D}{3-q_V}}}
{\left\{1+(q_V-1)\frac{(\Delta x_t)^2}{[T^V_{q_V}(t)]^{\frac{2}{3-q_V}}}
\right\}^{\frac{1}{q_V-1}+\frac{D-1}{2}}}\ (\forall q_V,\ \forall_D)
\end{equation}
The second moment of this distribution diverges for $q_V\geq 5/3$,
and the distribution becomes not normalizable for $q_V\geq 3$.

There is no particular reason for $T^V_{q_V}$ being equal to
$T^A_{q_A}$ but, following \cite{zu}, we shall use here the simplest
choice, i.e., $T^A_{q_A}(t)=T^V_{q_V}(t),\ \forall t$ (given by Eq.
(14)). We can now summarize the whole algorithm for finding a {\it
global} minimum of a given energy/cost function $E(\vec{x})$:
\begin{description}
\item
[(i)] Fix $(q_A,q_V)$. Start, at $t=1$, with an arbitrary value
$\vec{x}_1$ and a high enough value for $T_{q_V}(1)$ (say about 2 times
the height of the highest expected ``hill''of $E(\vec{x})$, and
calculate $E(\vec{x}_1)$;
\item
[(ii)] Then randomly generate $\vec{x}_{t+1}$ from $\vec{x}_t$
according to Eq. (21) to determine the {\it size} of the jump
$\Delta x_t$, and isotropically determine its {\it direction};
\item
[(iii)] Then calculate $E(\vec{x}_{t+1})$: \\
If $E(\vec{x}_{t+1})<E(\vec{x}_t)$, replace $\vec{x}_t$ by
$\vec{x}_{t+1}$; \\
If $E(\vec{x}_{t+1})\geq E(\vec{x}_t)$, run a random number $r\in [0,1]$: if
$r>P_{q_A}$ given by Eq. (5) with $T^A_{q_A}(t)=T^V_{q_V}(t)$,
retain $\vec{x}_t$; otherwise, replace $\vec{x}_t$ by
$\vec{x}_{t+1}$;
\item
[(iv)] Calculate the new temperature $T^V_{q_V}$ using Eq. (14) and
go back to (ii) until the minimum of $E(\vec{x})$ is reached within
the desired precision.
\end{description}

\section{A SIMPLE D=1 ILLUSTRATION}

In this Section we numerically treat a simple $D=1$ example with a
double purpose: on one hand to exhibit how the procedure works and,
on the other, to find for which pair $(q_V,q_A)$ the algorithm is the
{\it quickest}. (We recall that $(q_V,q_A)=(1,1)$ corresponds to CSA
and (2,1) to FSA).

We choose the same example treated in \cite{zu}, namely
\begin{equation}
E(x)=x^4-16x^2+5x+E_0
\end{equation}
where we have introduced the additive constant $E_0\simeq 78.3323$ so
that $E(x)\geq 0,\ \forall x$, thus satisfying the convention
adopted below Eq. (5); see Fig. 1. As initial conditions for all of
our runs we used $x_1=2$ and $T_{q_V}=100$. In Fig. 2 we can see
typical runs for $q_A=1.1$ and different values of $q_V$. Clearly the
case $q_V=2.5$ is much faster and precise than classical and fast
annealings ($q_V=1.1\simeq 1$ and $2$ respectively). To study the
$(q_V,q_A)$ influence on the quickness of the algorithm we have
adopted once for ever, an arbitrary convergence criterium. For each
$(q_V,q_A)$ pair we evaluate the mean value of $x_t$ in intervals of
$100$ time steps. Whenever two successive intervals presented mean values
whose difference was smaller than a precision $\varepsilon =10^3$, we
stopped the run. We then evaluated the total iteration time $\tau$
and repeated the whole annealing procedure $10$ times. Finally, we
compute the average total calculation time $<\tau >$. The
$(q_V,q_A)$ dependence of the average $<\tau >$ is presented in Fig.
3 for typical values of $q_A$.  Fig. 3 indicates
that machines with $q_A=1.1$ and $q_V=2.9$ are typically $5$ times
faster than the Cauchy machine \cite{zu}, which is in turn about
$5$ times faster than the Boltzmann machine \cite{ki,ge,zu}. We
have done our simulations on a 486 DX microcomputer with a clock
of 50 MHz. In this machine each one of the 10 solutions
demanded, approximately,
1 minute and 20 seconds of CPU time for $q_{V}=2$, and only 20
seconds for $q_{V}=2.5$. Finally
in Fig. 4 we present the dependence of $<\tau >$ with $q_A$ for
$q_V=2$; for this case the Cauchy machine \cite{zu} is the best
performant. These results indicate that the
quickest algorithm occurs for $q_A=1$ and $q_V=3$.

\section{CONCLUSION}
Inspired in a recent generalization of
Boltzmann-Gibbs statistical mechanics, we have heuristically
developed a {\it generalized simulated annealing} (characterized by the
parameters $(q_V,q_A)$) which unifies the so called {\it classical}
(Boltzmann machine; $q_V=q_A=1)$ and {\it fast} (Cauchy machine;
$q_V/2=q_A=1)$ ones. This computational method is based on
stochastics dynamics (which asymptotically becomes, as time runs to
infinity, a gradient descent method), and enables, with probability
one, the identification of a {\it global} minimum of any
(sufficiently nonsingular) given energy/cost function which depends
on a continuous D-dimensional variable $\vec{x}$. While the discrete
time $t$ increases, it might happen that $\vec{x}_t$ provisionally
stabilizes on a given value, and eventually abandons it running
towards the {\it global} minimum. This temporary residence can be
used, as bonus of the present method, to identify {\it some} of the
{\it local} minima. If sufficiently many computational runs are done
by starting at random initial positions $\{\vec{x}_1\}$, this
property could in principle be used to identify {\it all} the {\it
local} minima as well as {\it all} the {\it global} ones.

For simplicity, we have mainly discussed herein the restricted region
$q_V\geq 1$ and $q_A\geq 1$ (with $E(\vec{x})\geq 0,\ \forall
\vec{x}$), and have identified the $(q_V,q_A)\simeq (2.9,1)$ machines
as the most performant ones in practical terms. This algorithm has
been illustrated herein with a simple two-minima $D=1$ energy
function, and has already been successfully used \cite{mun}
for recovering the global energy minima (with respect to the
dihedral angle) of a variety of simple molecules (e.g., $CH_3OH,\ H_2O_2,\
C_2H_6$). It should be very interesting to test the present generalized
algorithm with many-body systems presenting a great number of minima
(spin-glass-like frustrated systems, traveling salesman problem , neural
networks, complex economic systems).

We acknowledge N. Caticha for stressing our attention onto Szu's
algorithm \cite{zu}, as well as K.C. Mundim and A.M.C. de Souza for very
useful discussions. At the time of submission of this paper we
took notice that this algorithm (informally communicated by us
to a few people) has already been succesfully implemented for
the Traveling Salesman Problem \cite{thadeu1,pablo}, for fitting
curves \cite{thadeu2,thadeu3}, and for a problem
of genetic mutations \cite{miron}.

\newpage
\section*{Caption for figures}

\begin{description}
\item
[Fig. 1:] $D=1$ energy function $E(x)=x^4-16x^2+5x+E_0$ vs. $x$ for
$E_0=78.3323$; global minimum
$x_1^{\ast}= -2.90353$ and $E(x_1^{\ast})=0$; maximum:
$x_2^{\ast}=0.156731$ and $E(x_2^{\ast})=78.7235$; local
minimum: $x_3^{\ast}=2.74680$ and $E(x_3^{\ast})=28.2735$.
\item
[Fig. 2:] Typical runs of the GSA algorithm $x_t$ vs. $t$ (annealing
time) for initial conditions $x_1=2$, $T_{q_V}(1)=100$. All four runs
correspond to $q_A=1.1$; a) $q_V=1.1$, b) $q_V=1.5$, c) $q_V=2$, d)
$q_V=2.5$. Notice the different scales for the ordinates.
\item
[Fig. 3:] Average total calculating time $<\tau >$ vs. $q_V$ for two
typical values of $q_A$ (${\bf \triangle}$: $q_A=1.5$; $\odot$: $q_A=1.1$).
\item
[Fig. 4:] Average total calculation time $<\tau >$ vs. $q_A$ for $q_V=2$.
\end{description}

\end{document}